\documentclass{aa}
\usepackage{graphics}
\begin{document}

\thesaurus{03(09.04.1:11.16.1:11.19.2:11.19.6)}

\title{Is the exponential distribution a good approximation of dusty galactic
disks?}

   \author{A. Misiriotis,\inst{1,2} 
          N.D. Kylafis,\inst{1,2}
          J. Papamastorakis,\inst{1,2} and
          E.M. Xilouris\inst{2,3}}

   \authorrunning{A. Misiriotis et. al.}

   \offprints{angmis@physics.uch.gr}

   \institute{University of Crete, Physics Department, P.O. Box 2208, 710 03
   Heraklion, Crete, Greece
   \and Foundation for Research and Technology-Hellas, P.O. Box 1527, 711 10 
   Heraklion, Crete, Greece
   \and University of Athens, Department of Physics, Section of Astrophysics, 
   Astronomy \& Mechanics, 157 83 Athens, Greece
   \and Observatoire de Marseille, 2 Place Le Verrier 13248 Marseille
   Cedex 4, France}

   \date{Received \ \ \ \ \ \ \ \ ; Accepted }

   \maketitle

\begin{abstract}
We investigate how significant the spiral structure is on calculations 
concerning radiative transfer in dusty spiral galaxies seen edge-on.
The widely adopted exponential disk model (i.e. both the stars and the dust
are distributed exponentially in the radial direction and also perpendicular 
to the plane of the disk) is now subject to a detailed 
comparison with a realistic model that includes spiral structure
for the stars and the dust in the disk. In particular, model images
of galaxies with logarithmic spiral arms are constructed, such that
the azimuthally averaged disk is exponential in radius and in height,
as the observations suggest.
Then, pure exponential disk models (i.e. with no spiral structure)
are used to fit the edge-on appearance of the model images.
As a result, the parameters derived after the fit are compared
to the real values used to create the spiral-structured images.
It turns out that the
plain exponential disk model is able to give a very good description
of the galactic disk with its parameters varying only by a few
percent from their true values.
\end{abstract}


\section{Introduction}
Modeling the dust and stellar content of spiral galaxies
is a very crucial procedure needed for the correct interpretation
of the observations. The amount of interstellar dust embedded inside
spiral galaxies, the way that dust is distributed within spiral galaxies
and also the
extinction effects of the dust to the starlight are some of the
questions that can be answered by performing radiative transfer 
modeling of individual spiral galaxies. 

One very important thing
that needs consideration when doing such analysis is the right
choice of the stellar and dust distributions. 
In particular, the galactic disk is a quite complex system,
where stars and dust are mixed together usually in a spiral formation.
For this reason, one has to use realistic distributions able
to reproduce quite accurately the observations. On the other hand,
simple mathematical expressions for these distributions are chosen
in order to keep the free parameters to the minimum.

For the distribution of the starlight in the disk of spiral galaxies,
the exponential function is very widely in use. This simple
mathematical expression is able to describe
the distribution of stars in both directions, radially and perpendicular to the
disk. Decomposition techniques used by different authors in order
to separate the bulge and the disk component strongly support
this argument. For galaxies seen face-on (and at moderate inclination
angles), radial profile
fitting (e.g. Freeman 1970), fitting to azimuthally averaged
profiles (e.g. Boroson 1981), as well as ellipse fitting techniques to 2D images
(e.g. de Jong 1995) show that the exponential in the radial  distance $R$ 
is a good representation of galactic disks with only
small deviations mainly due to the spiral structure of the galaxy
(see Serna 1997).
Other works like those of Shaw \& Gilmore (1989) and de Grijs (1997) dealing 
with modeling of edge-on galaxies support the idea that
exponential functions are good representations also for the
$z$ (vertical to the disk) direction. 

Performing radiative transfer modeling of edge-on galaxies,
Xilouris et al. (1997, 1998, 1999) found that exponential functions for
the luminosity density of the stars in the disk as well as for the
extinction coefficient give an excellent description of the observations.
The advantage of modeling galaxies in the edge-on configuration is that
the integration of light along the line of sight is able to cancel
out most of the structure of the galaxies (i.e. spiral structure)
and therfore allows for simple functions such as exponentials to 
give good representation of the observations. Thus, although in the
face-on configuration a large variation between arm and interarm regions 
might be present for both the stars and the dust 
(White \& Keel 1992, Corradi et al. 1996, Beckman et al. 1996,
Gonzalez et al. 1998), in the edge-on case an average description
of the galaxy characteristics can be obtained quite accurately.
We are going to investigate the validity of this argument 
by comparing the exponential distributions
with more realistic distributions which
include spiral structure.

In Sect. 2 we describe the method that we use to address this problem
and in Sect. 3 we present the results of our calculations.
Finally, in Sect. 4 we summarize our work.


\section{Method}
\begin {figure*}[!t]                                                           
\resizebox{\hsize}{!}{\includegraphics{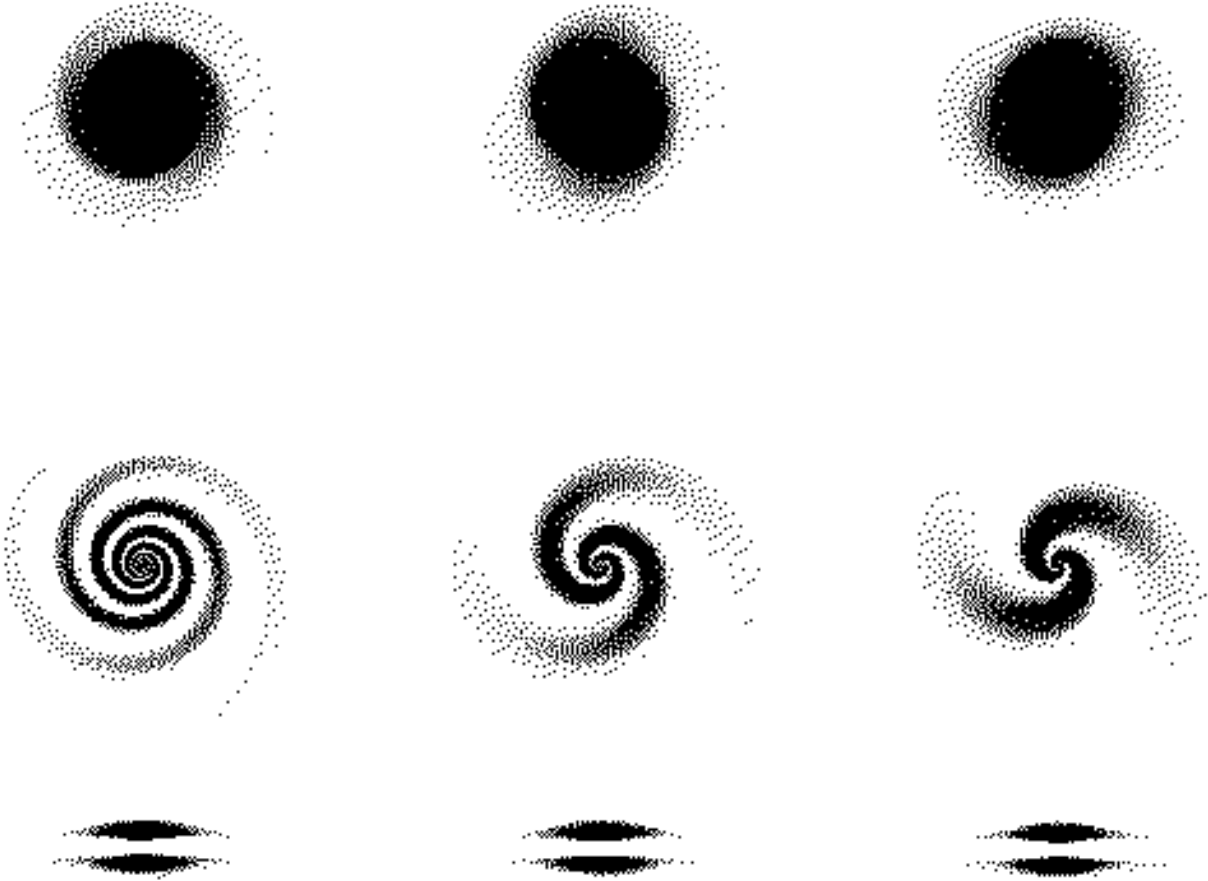}}                               
\caption{Face-on surface brightness of model galaxies (top panel) with
different values of the pitch angle ($10\degr, 20\degr$ and $30\degr$          
from left to right). The distribution of the face-on
optical depth for these galaxies   
is shown in the middle, while their edge-on appearance                           
(looking from $\phi = 0\degr$)                                                  
is shown in the bottom panel.}                                                  
\label{fig1}                                                                    
\end{figure*}    

The method that we follow in this work consists of two basic steps.
In the first step, model galaxies with realistic spiral structure
are constructed. After a visual inspection of the face-on appearance of
these models to see the spiral pattern, we create their 
edge-on images which are now treated as real observations. In the 
second step we fit these ``observations'' with a galaxy model
where now the galactic disk is described by the widely used plain
exponential model. In this way, a comparison between the parameters
derived from the fitting procedure and those 
used to produce the artificial
``observations'' can be made and thus a quantitative answer
about the validity of the plain exponential model as an
approximation to galactic disks can be given.

\subsection{Artificial spiral galaxies}

We adopt a simple, yet realistic, distribution of stars and dust in the 
artificial galaxy.  A simple expression is needed in order to keep the number 
of free parameters as small as possible and thus have a better
control on the problem.  A realistic spiral structure is that of logarithmic 
spiral arms (Binney \& Merrifield 1998).  
Thus, a simple but realistic artificial 
spiral galaxy is constructed by imposing the logarithmic spiral arms as a 
perturbation on an exponential disk.  In this way, the azimuthally averaged
face-on profile of the artificial galaxy has an exponential radial 
distribution. 

For the stellar emissivity we use the formula 
\begin{displaymath}
L(R,z) = L_s \exp \left( - \frac{R}{h_s} - \frac{|z|}{z_s} \right)
\end{displaymath}
\begin{displaymath}
~~~~~~~~~~\times \left\{1+w_s \sin\left[\frac{m}{\tan(p)}\log(R) - m\phi \right]\right\}
\end{displaymath}
\begin{equation}
~~~~~~~~~~+L_b \exp (-7.67 B^{1/4}) B^{-7/8}~.
\end{equation}
In this expression the first part describes an
exponential disk, the second part gives the spiral perturbation and
the third part describes the bulge, which in projection is the well-known
$R^{1/4}$-law (Christensen 1990).
Here $R$, $z$ and $\phi$ are the 
cylindrical coordinates, $L_s$ is the stellar emissivity per unit volume
at the center of the disk and $h_s$ and $z_s$ are the scalelength and 
scaleheight respectively of the stars in the disk.

The amplitude of the spiral perturbation is described by the parameter
$w_s$. When $w_s=0$ the plain exponential disk is obtained, while
the spiral perturbation becomes higher with larger values of $w_s$.
Another parameter that defines the shape of the spiral arms is
the pitch angle $p$. Small values of $p$ mean that the
spiral arms are tightly wound, while larger values produce a 
looser spiral structure. The integer $m$ gives the number
of the spiral arms.

For the bulge, $L_b$ is the stellar emissivity per unit volume
at the center, while $B$ is defined by
\begin{equation}
B = \frac{\sqrt{R^2 + z^2 (a/b)^2}}{R_e} ,
\end{equation}
with $R_e$ being the effective radius of the bulge and $a$ and $b$ being the
semi-major and semi-minor axis respectively of the bulge.

For the dust distribution we use a similar formula as that
adopted for the stellar distribution in the disk, namely
\begin{displaymath}
\kappa(R,z) = \kappa_{\lambda} \exp \left( - \frac{R}{h_d}
- \frac{|z|}{z_d} \right)
\end{displaymath}
\begin{equation}
~~~~~~~~~~\times \left\{1+w_d \sin\left[\frac{m}{\tan(p)}\log(R) - m\phi \right]\right\},
\end{equation}
where $\kappa_{\lambda}$ is the extinction coefficient at
wavelength $\lambda$ at the center of the disk and
$h_d$ and $z_d$ are the scalelength and scaleheight respectively
of the dust. Here $w_d$ gives the amplitude of the spiral 
perturbation of the dust.  Note that the angle $\phi$ here need not be the same
as that in Eq. (1). The stellar arm and the dust arm may have a phase
difference between them.

\begin{table}
\caption[]{Parameters used to describe a typical spiral galaxy.}
\begin{tabular}{llllllll}
\hline
\multicolumn{1}{c}{Parameter} &
\multicolumn{1}{c}{Units} &
\multicolumn{2}{c}{B band}\\
\hline
$I_s$              & mag/arcsec$^2$ & $20.0$ \\
$z_s$              & kpc            & $0.4$  \\
$h_s$              & kpc            & $5.0$  \\
$I_b$              & mag/arcsec$^2$ & $12.0$ \\
$R_e$              & kpc            & $1.5$  \\
$b/a$              & --             & $0.5$  \\
$\tau_{\lambda}^e$ & --             & $27$   \\
$z_d$              & kpc            & $0.2$  \\
$h_d$              & kpc            & $6.3$  \\
$w_d$              & --             & $0.4$  \\
$w_s$              & --             & $0.3$  \\
$m$                & --             & $2$    \\
$p$                & degrees        & $10,20,30$ \\
\hline
\end{tabular}
\end{table}

\begin {figure}[!t]                                                             
\resizebox{\hsize}{!}{\includegraphics{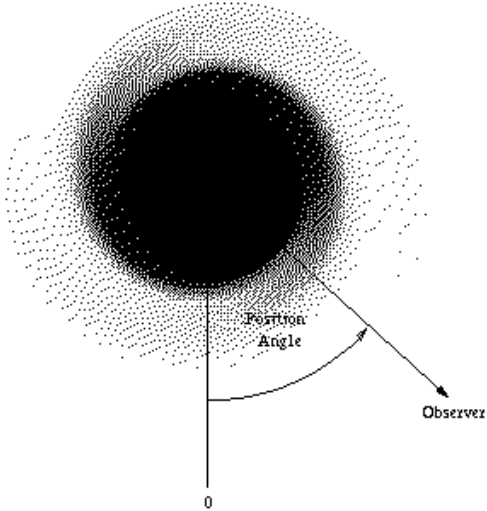}}                               
\caption{Schematic representation of the position angle. 
The value of $0\degr$               
is as shown in the figure, while the position angle
increases when the line of sight moves counterclockwise.} 
\label{fig2}                                                                    
\end{figure}                                                                    

\begin {figure}[!t]
\resizebox{\hsize}{!}{\includegraphics{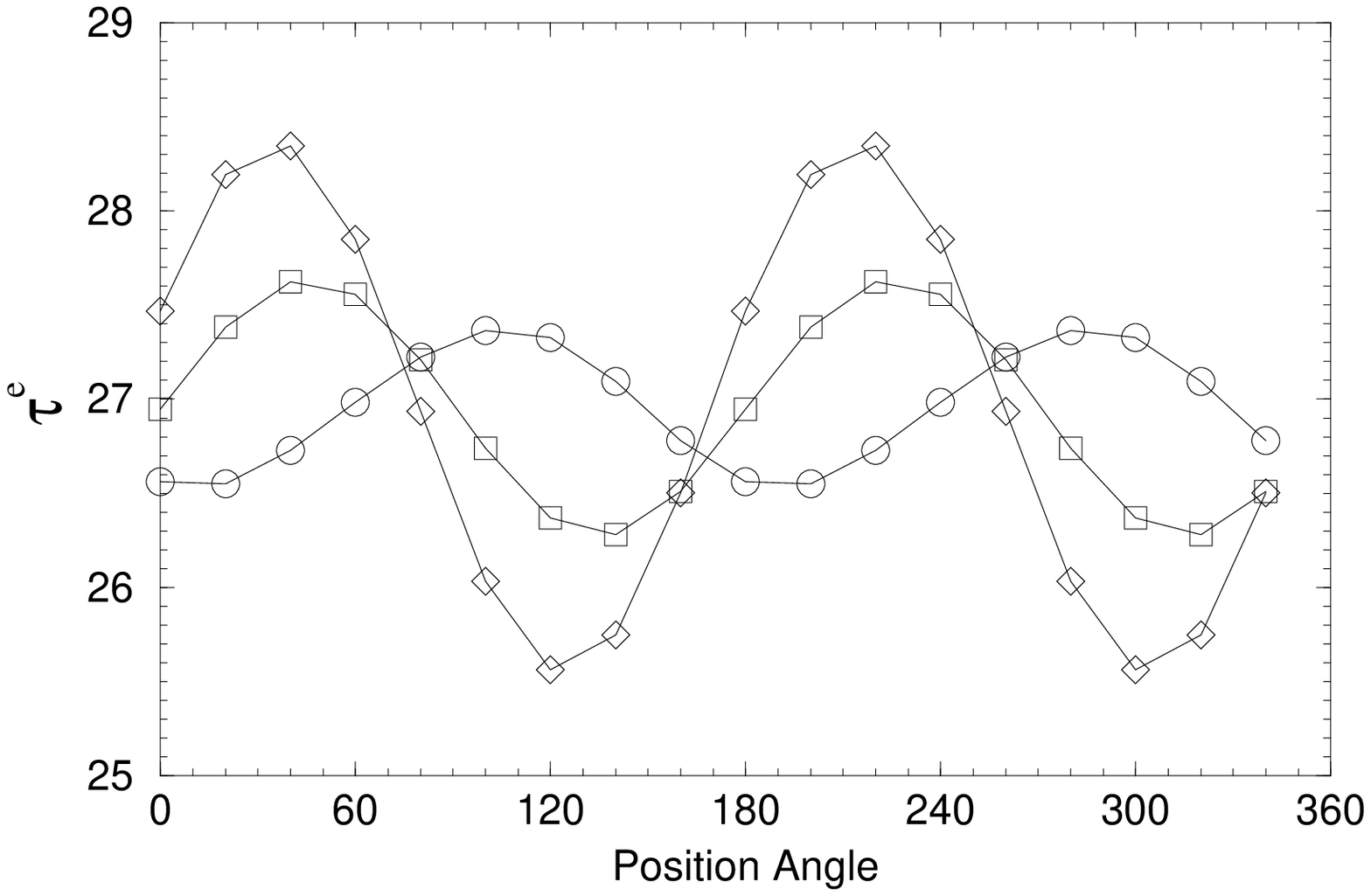}}                               
\caption{The central edge-on optical depth of the model spiral galaxy as        
a function of position angle. Circles, squares and diamonds refer to 
three different models with pitch angle $10\degr$, $20\degr$ and        
$30\degr$ respectively.}
\label{fig3}                                                                    
\end{figure}

For the parameters describing the exponential disk of the stars
and the dust as well as the bulge characteristics we use the 
mean values derived from the B-band modeling of seven
spiral galaxies presented in Xilouris et al. (1999).

Since the most dominant spiral structure in galaxies 
is that of the two spiral arms (Kennicutt 1981;
Considere \& Athanassoula 1988; Puerari \& Dottori 1992) 
we only consider models
where $m=2$.
Galaxies with strong one-arm structure do exist, but they
constitute a minority (Rudnick \& Rix 1998).

For the parameter $w_d$ we take the value of 0.4. 
With this value the optical depth calculated in the arm region
is roughly twice as much as in the inter-arm region. This is in
good agreement with studies of overlapping galaxies (e.g. 
White \& Keel 1992).

For $w_s$ we use the value of 0.3 resulting (with the 
extinction effects included)
in a spiral arm amplitude of $\sim 0.1 - 0.2$ mag, which is a typical
amplitude seen in radial profiles of face-on spiral galaxies and
reproduces the desired strength for the spiral arms (Rix \& Zaritsky 1995). 

For the pitch angle $p$ we consider the cases of 
$10\degr, 20\degr$ and $30\degr$,
which give a wide variety of spiral patterns from tightly wound to loosely
wound.  All the parameters mentioned above are summarized in Table 1.

The radiative transfer is done in the way described by Kylafis
\& Bahcall (1987; see also Xilouris et al. 1997). 
As described in detail in these references, 
the radiative transfer code is capable of dealing with both
absorption and scattering of light by the interstellar dust and
also of allowing for various distributions for the stars and the dust.

Using the model described above and the parameters given in Table 1
we produce the images shown in Fig. 1. The top three panels of this figure
show the face-on surface brightness distribution of such a galaxy for the
three values of the pitch angle ($10\degr, 20\degr$ and $30\degr$) from
left to right. The spiral structure is evident in these images
with the spiral arms being more tightly wound for $p = 10\degr$
and looser when $p = 30\degr$.

In the middle three panels of Fig. 1 we show the distribution of the
optical depth when the galaxy is seen face-on for the three different
values of the pitch angle mentioned above. In these pictures one can 
follow the spiral pattern all the way to the center of the galaxy since
the bulge is assumed to contain no dust.

Finally, in the last three panels of Fig. 1 one can see the corresponding
edge-on appearance of the galaxies shown face-on in the top three panels.

One thing that is very obvious from Fig. 1 (top and middle panels)
is that the galaxy is no more axisymmetric as it is the case
in the plain exponential disk model.
The spiral structure that is now embedded in the model as a
perturbation in the disk has broken this symmetry. Thus, in order to 
do a full analysis of the problem we have to examine
the galaxy from different azimuthal views (position angles).
To do so we have created nine edge-on model galaxies (for each
of the three different pitch angles considered here), covering 
the range from $0\degr$ to $160\degr$ with
a step of $20\degr$ for the position angle.  
For the definition of the position angle, see Fig. 2. 
Since the galaxy has exactly the same appearance in the interval from
$180\degr$ to $360\degr$, we only consider the range of position angles mentioned above.

To demonstrate this asymmetry more 
quantitatively we have computed
the central edge-on optical depth ($\tau^e$) for all these nine model galaxies.
Unlike the plain exponential disk model where $\tau^e$ can be calculated
analytically ($\tau^e = 2 \kappa_{\lambda} h_d$), here we have to perform 
numerical integration of Eq. (3) along the line of sight that passes
through the center. The value of $\tau^e$
is shown in Fig. 3 as a function of the position angle. In order to have the 
full coverage in position angle (from $0\degr$ to $360\degr$), the
values calculated in the interval ($0\degr$ - $180\degr$) were
repeated in the interval ($180\degr$ - $360\degr$). In this figure, 
the three models constructed
with pitch angles $10\degr, 20\degr$ and $30\degr$ 
are denoted with circles, squares
and diamonds respectively. In all three cases, a variation of the optical depth
with position angle is evident. The largest variation is
found for the case where the pitch angle is $30\degr$ and
it is $\sim 5\%$. It is obvious that all the values are around
the true value of 27, used to construct the galaxy (see Table 1).

\subsection{The fitting procedure}
The edge-on images created as described earlier are
now treated as ``observations'' and with a fitting procedure we
seek for the values of the parameters of the plain exponential disk
that gives the best possible representation of the ``observations''.
The fitting
algorithm is a modification of the Levenberg-Marquardt routine
taken from the Minipack
library. The whole procedure is described in detail in Xilouris et al.
(1997). 

Preliminary tests have shown that the derived values of the parameters
 describing the bulge
are essentially identical to the real values used to construct the
model images. In order to simplify the fitting process and since we are
only interested in the disk, the bulge parameters were kept constant
during the fit. Six parameters are now free to vary. These are
the scalelength and scaleheight of the stellar disk with its
central surface brightness ($h_s$, $z_s$ and $I_s$ respectively)
as well as the scalelength and scaleheight of the dust and the
central edge-on optical depth ($h_d$, $z_d$ and $\tau^e$ respectively).

\section{Results}

\begin {figure*}[!ht]
\resizebox{\hsize}{!}{\includegraphics{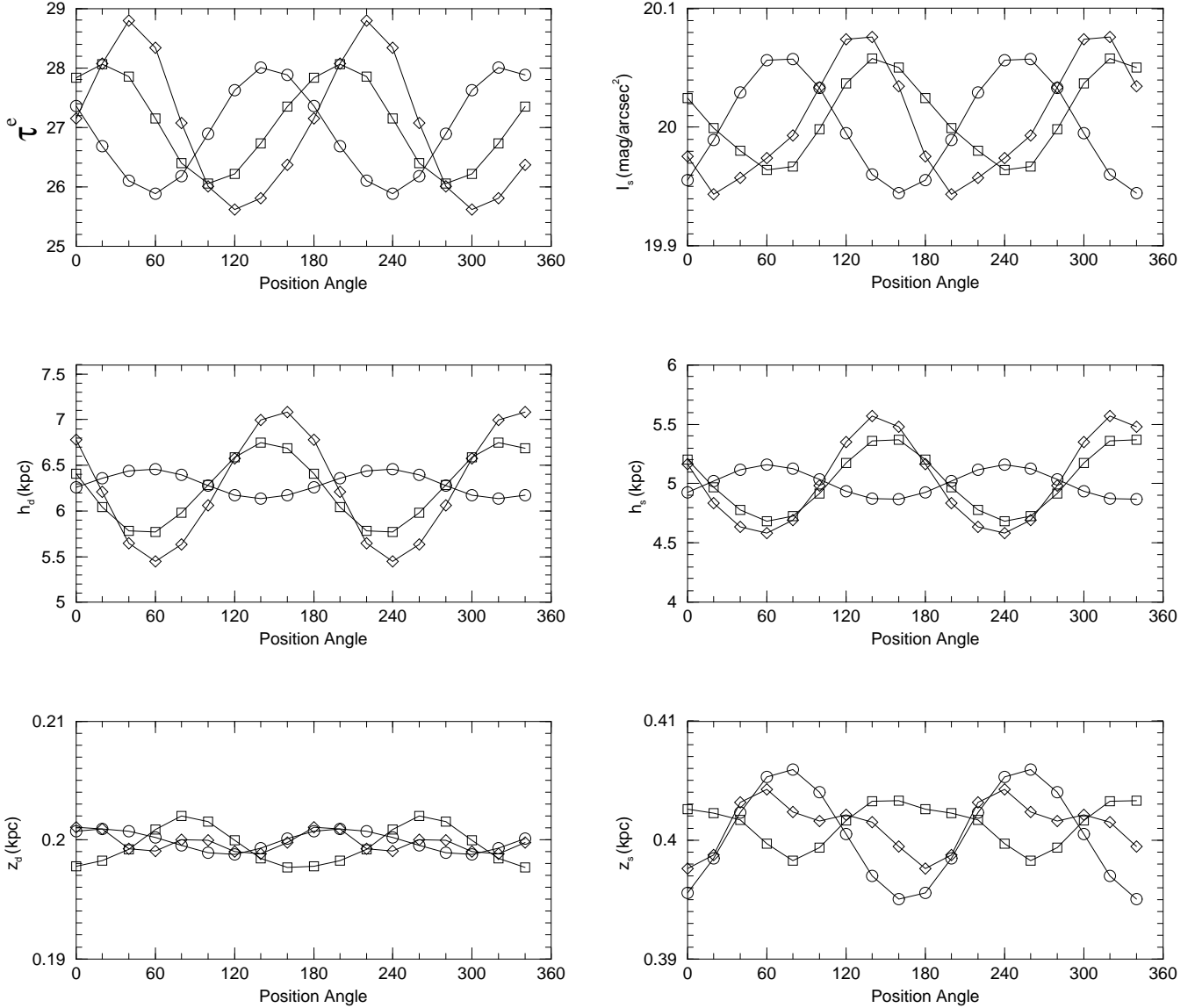}}
\caption{Disk parameters deduced by fitting the exponential model to the
images of our artificial galaxy for several position angles around
the galaxy. In all these sets of images the dust is in phase with the stars.
Circles, squares and diamonds refer to 
three different models with pitch angle $10\degr$, $20\degr$ and        
$30\degr$ respectively.                                                        
The top left graph shows the variation
of the deduced optical depth. The top right graph shows the variation of the
inferred central intensity of the disk. The middle left graph shows the 
variation of the derived
dust scalelength, while the middle right graph shows that 
of the stellar
scalelength. 
Finally, the bottom left graph shows the variation of the scaleheight of
the dust and the bottom right graph the variation of the scaleheight of the
stars.
}
\label{fig4}
\end{figure*}

\begin {figure*}[!ht]
\resizebox{\hsize}{!}{\includegraphics{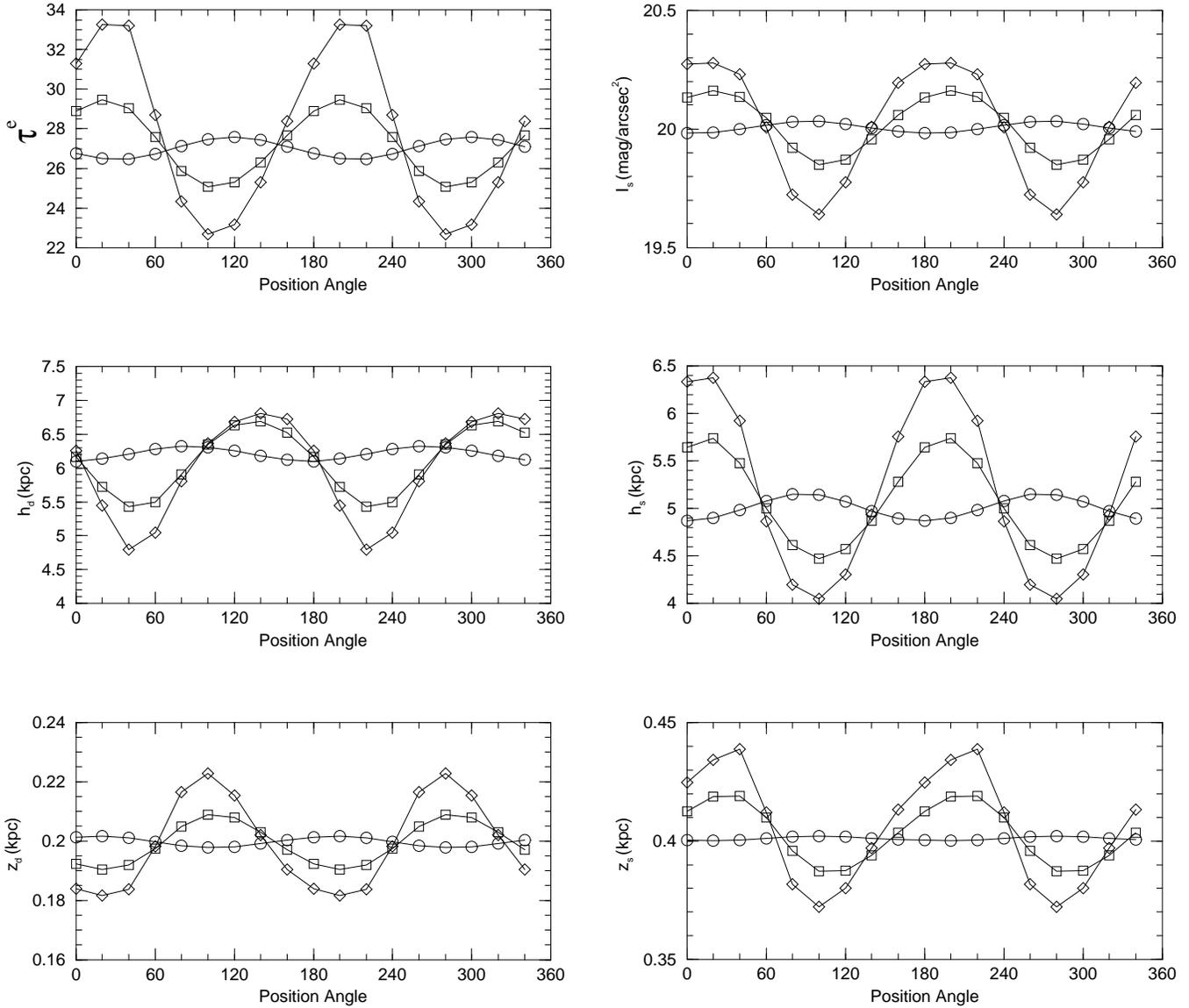}}
\caption{The same as in Fig. 4, but the stellar arms are leading the dust arms
by $30\degr$.}
\label{fig5}
\end{figure*}

Figure 4 shows six graphs. The top left graph gives the variation of the
deduced edge-on optical depth of the galaxy as we observe it from different 
angles. From this graph one can see that the variation of the deduced optical
depth from different points of view is no more than 6\% different than
 the mean value
of the optical depth. Furthermore, a comparison with Fig. 3, which shows the 
real value of the optical depth, reveals that there is no systematic error
in the derived value. The deviations are equally distributed around
$\tau^e=27$, which is what we would have without the spiral structure.
The variation is of
the same order of magnitude regardless of the pitch angle. The top right 
graph in Fig. 4 presents the deduced central luminosity of the disk.
This graph shows that the variation of the inferred central luminosity of the
stellar disk is very small and it is weakly dependent on the pitch angle.
In the middle left graph of Fig. 4 the derived scalelength of the dust 
is presented. The variation of the derived value is
about 5\% for the 10-degree pitch angle and goes up to 17\% for
the 30-degree pitch angle. The increase of the variation with increasing
pitch angle is expected, because for large pitch angles the spiral arms
are loosely wound, thus causing the galaxy to be less axisymmetric.
As a result, from some
points of view the dust seems to be more concentrated to the center of the 
galaxy and from other points of view more extended. 

In the middle right graph
of Fig. 4, that shows the scalelength of the stars, it is evident that
the same effect occurs for the stars as well. Certain points of view give the
impression of a more centrally condensed disk, while others of an 
extended disk.  The fact that we have taken
the stellar and the dust spiral structure to be in
phase (i.e. the dust 
spiral arms are neither trailing nor leading the stellar spiral arms)
causes the deduced 
scalelengths of the dust and the stars to also vary in phase. The case of a 
phase difference is examined below.

The bottom left and bottom right graphs of Fig. 4 show the 
variation of the scaleheight of the dust and the stars respectively. The 
variation of both scaleheights is negligible. 
This is an attribute of the formula
we used for our artificial galaxy. Since the spiral variation we added to the
exponential disks is not a function of $z$
it is expected that in the $z$ direction
our artificial galaxy behaves exactly as the exponential model.

There are indications
(van der Kruit \& Searle 1981;Wainscoat et al. 1989) that the
dust arms are not located exactly on the stellar arms. Thus, we re-created the 
edge-on images, but this time the stellar spiral arms were set to 
lead the dust arms
by 30 degrees. We then fitted the new images with the exponential model 
and the deduced parameters are shown in Fig. 5. 
As in Fig.4,  the top left graph of Fig. 5
shows the optical depth as a function of position angle. A comparison of
this graph with the corresponding graph in Fig. 4 reveals that the variation
of the values derived from the new set of images is significantly larger. 
The origin of 
this effect is the fact that the dust is either in front of the stars (for some
position angles) or behind the stars (for other position angles). A strong
dependence on the pitch angle is also evident.  Note,
however, that the mean of all the derived values is unaffected. The same
effect can be seen on the top right graph of Fig. 5, where the derived 
central luminosity of the disk is plotted. The variation of the central
luminosity is again larger than in the previous case, but the mean value is
equal to the true one. 

In the middle left graph of Fig. 5 we show the 
scalelength of the dust as a function of the position angle. The variation
of the derived dust scalelength can differ as much as 25\% for a galaxy
with pitch angle equal to 30 degrees. But again the mean value for all 
position angles is identical to
the one we used to create the images. In the 
middle right graph one can see that the scalelength of the stars also varies
as much as 30\%,but the mean of all the derived values is
the correct one. 

The left and right bottom graphs show the variation of the 
scaleheights of the dust and the stars, which is about 10\% for the worst
case of a 30-degree pitch angle and is practically zero for a galaxy
with more tight arms. 

The most important conclusion of all the graphs is
that the derived values of all quantities tend to distribute equally
around the real value we used to create the artificial images.

\section{Summary}
In our attempt to investigate how significant the spiral structure is 
when doing radiative transfer modeling of spiral galaxies seen edge-on,
we constructed a model galaxy with very prominent spiral arms in the disk.
This quite realistic image of the galaxy is now treated as observation and
the widely adopted exponential model for the galactic disk is now
fitted to the data. 

This analysis shows that the plain exponential
disk model is a very accurate description for galactic disks seen edge-on
with only small deviations of its parameters from the real ones (typically
a few percent). Furthermore, the variation from the real parameters 
would be averaged out if we could see the same galaxy from 
several point of views.
This is of course impossible for an individual galaxy,
but suggests that if the exponential
model is used for a statistical study of many edge-on galaxies no systematic
error is introduced. Thus, we conclude that the exponential model is a very 
good approximation of the galactic disks.

\end{document}